\documentclass[12pt]{article}

 \usepackage[utf8]{inputenc}

\usepackage{times}

\usepackage{amsmath}
\usepackage{amsfonts}
\usepackage{amssymb}
\usepackage{graphicx}
\usepackage{booktabs}

\newenvironment{sciabstract}
{ \begin{quote} \bf }
{ \end{quote} }

\title{The Network of U.S. Mutual Fund Investments: Diversification, Similarity and Fragility throughout the Global Financial Crisis}

\author{
  Danilo Delpini,${}^{1,2\ast}$ Stefano Battiston,${}^{3}$ Guido Caldarelli,${}^{2,4}$ Massimo Riccaboni${}^{2,5}$\\
\\
\normalsize{${}^{1}$Dept of Economics and Business, University of Sassari, Sassari, Italy}\\
\normalsize{${}^{2}$IMT Institute for Advanced Studies, Lucca, Italy}\\
\normalsize{${}^{3}$Dept of Banking and Finance, University of Zurich, Zurich, Switzerland}\\
\normalsize{${}^{4}$ISC-CNR Uos ``Sapienza'', Roma, Italy}\\
\normalsize{${}^{5}$Dept of Managerial Economics, Strategy and Innovation, Katholieke Universiteit Leuven, Leuven, Belgium}\\
\\
\normalsize{$^\ast$E-mail: ddelpini@uniss.it}
}

\date{}

\begin{document} 


\baselineskip24pt


\maketitle



\begin{sciabstract}
Network theory proved recently to be useful in the quantification of many properties of financial systems.
The analysis of the structure of investment portfolios is a major application since their eventual correlation and overlap impact the actual risk diversification by individual investors.
We investigate the bipartite network of US mutual fund portfolios and their assets.
We follow its evolution during the Global Financial Crisis and analyse the interplay between diversification, as understood in classical portfolio theory, and similarity of the investments of different funds.
We show that, on average, portfolios have become more diversified and less similar during the crisis.
However, we also find that large overlap is far more likely than expected from models of random allocation of investments.
This indicates the existence of strong correlations between fund portfolio strategies.
We introduce a simplified model of propagation of financial shocks, that we exploit to show that a systemic risk component origins from the similarity of portfolios.
The network is still vulnerable after crisis because of this effect, despite the increase in the diversification of portfolios.
Our results indicate that diversification may even increase systemic risk when funds diversify in the same way.
Diversification and similarity can play antagonistic roles and the trade-off between the two should be taken into account to properly assess systemic risk.
\end{sciabstract}


\section*{Introduction}

One of the most important issues that the Global Financial Crisis (GFC) of 2007-2008 has highlighted concerns the effects on systemic risk of the increasing interdependence both between large institutional investors and among global assets.
On the one hand, large institutional investors allow for a better diversification of individual risk: the larger the number of 
different assets in a portfolio, the smaller the fraction of an idiosyncratic shock an investor has to bear. 
On the other hand, the GFC has shown that cross-sectional dependencies between assets can cause idiosyncratic shocks (i.e. related to the distress/bankruptcy  of a single specified asset) to spread, ultimately substantially threatening the stability of the entire financial system.
According to classical investment portfolio theory, risk depends on the share of individual stock holdings and the variance--covariance matrix among its 
holdings~\cite{Statman:1987aa}.
Hence, theoretical models imply that a portfolio should be fully diversified to reduce risks (unsystematic risks), but how to construct a 
well-diversified portfolio still remains to be studied.
There is consensus in the literature that a strong risk reduction of holdings can be realized by increasing the 
number of assets in a portfolio~\cite{Domian:2003aa,Domian:2007aa,Statman:2004aa}. Indeed, between 1997 and 2012, assets in the equity, balanced, and
fixed income mutual funds have increased by more than 400 percent~\cite{Sialm:2015aa}.
The efficacy of diversification strategies depends, however, on market conditions and systemic risk that may moderate the relationship between diversification, fund performance, and risk.
Recently, it has been shown that the benefit of diversification increases within higher market volatility conditions, such as the GFC of 2007-2008 meaning that the number of stocks needed to achieve a well-diversified portfolio increases under those market conditions~\cite{Hu:2014aa}. Unfortunately, the implications of such an increase of portfolio diversification during crises are not fully understood.
Moreover, the empirical evidence accumulated during the GFC has raised legitimate doubts on the effectiveness of portfolio diversification strategies 
to reduce risk.

Systemic risk in financial systems has been increasingly investigated through the lens of network theory~\cite{Battiston:2016ac,Battiston:2016ab,Acemoglu:2015aa,Battiston:2012ac,Battiston:2012ab,Elliott:2014aa,Galbiati:2013aa,Delpini:2013aa}.
Most of the work so far has focused on the network of interbank loans, even though there is a paucity of data about the real-world structure of financial networks. 
Interestingly, it has been found that, when the magnitude of negative shocks is large and the network is scale-free, a more densely 
connected financial network (corresponding to a more diversified pattern of investment) serves as a mechanism for the propagation of shocks, leading to a 
more fragile financial system, thus increasing systemic risk.

Depending on the level of similarities across mutual funds, increasing portfolio diversification during a crisis might increase the cross-correlation 
among assets thus amplifying systemic risk.
Therefore, the exact role played by the evolution of the financial network during crisis in creating systemic risk remains, at best, imperfectly understood. 
The same holds for the role of global fund managers which appears to have been studied mostly in simulated scenarios~\cite{Feldman:2010aa}.
So far, most of the papers in the literature assume that the relationship between diversification and risk is constant in the long run, 
regardless of market conditions. We contribute to fill this gap in the literature by analysing the bipartite network of US mutual funds 
over time and throughout the Global Financial Crisis.

Here, we provide an extensive characterization of the mutual fund universe covered by the CRSP Survivor-Bias-Free Mutual Funds Database in 2005-2010.
In particular, we consider the bipartite network of portfolio holdings and we perform a temporal analysis of its topology, focusing on the interplay 
between the complementary notions of diversification and similarity of portfolio investments.
We study how the systemic fragility of the system depends on the overlap between portfolios due to correlated investment strategies. This is done through simulations of a simple dynamics 
of distress propagation and through comparison with null models of random investments.
The fund holding network has been considered previously in~\cite{Wool:2013aa} where the correlation between changes in a firm's position in the network and future stock market performance is considered, and in~\cite{Abdesaken:2013aa} where the CRSP database was studied to detect possible conflicts of interest in the strategies of multi-fund managers.
In~\cite{Schwarzkopf:2010aa} the size distribution of funds has been investigated in detail, while in~\cite{Schwarzkopf:2016aa} a size-dependent model of fund growth has been proposed to explain its shape.
However, to the best of our knowledge, this is the first attempt to provide a comprehensive description of this system across years, 
from the point of view of network science, and to disentangle the roles of diversification and similarity in determining distress propagation and systemic fragility.

\section*{Results}

We analyse the mutual funds market as a bipartite network where one of the two sets is formed by the US mutual fund portfolios and the other set is made by the assets in those portfolios.
A detailed formalization is provided in Materials and Methods.
Within this framework, we perform an extensive statistical analysis of portfolio diversification from a twofold perspective.
On one side, we consider the usual notion of diversification \emph{within} a single portfolio, dating back to Markowitz~\cite{Markowitz:1952aa}.
This depends on the number of different assets in the portfolio, as well as on their weights, and on the correlation between asset price variations.
On the other side, we add a second dimension to the analysis, by studying diversification \emph{across} portfolios, defined as the degree of similarity 
of the assets held by mutual funds and measured by suitable overlap indices.
We compare the real data with two null models of random allocation of investments to test for heterogeneity and non trivial correlations between portfolios.
Finally, introducing a simple model of distress propagation and exploiting the random benchmarks, we explore the connection between similarity 
of portfolios, distress propagation and systemic fragility.

The above characterization is performed with special attention to the evolution of the network throughout the GFC.
In particular, we show how the systemic fragility of the network has changed. As it is well known, the peak of the crisis has been reached in 2008, 
culminating in the collapse of Lehman Brothers in September.
In order to compare the structure of the network over time, we select three \emph{reference quarters}, 2006Q2, 2008Q2, 2010Q2, where the notation ``2006Q2'', 
for instance, stands for the second quarter of year 2006. Our goal is to identify three moments in time \emph{before}, \emph{during} and
\emph{after} crisis.

\subsection*{Topology of fund holdings, concentration and similarity}
In this Section we investigate diversification within and across portfolios, through the lens of network science.
These concepts are related to the \emph{degree} of funds and assets in the holding network.

The degree $k_i$ of fund $i$ represents the number of different assets held at a given moment in time, while the degree $k_{\alpha}$ of an asset represents the number of funds that hold asset $\alpha$ in their portfolio.

\begin{figure}[!tb]
  \centering
  \includegraphics[width=1.\textwidth]{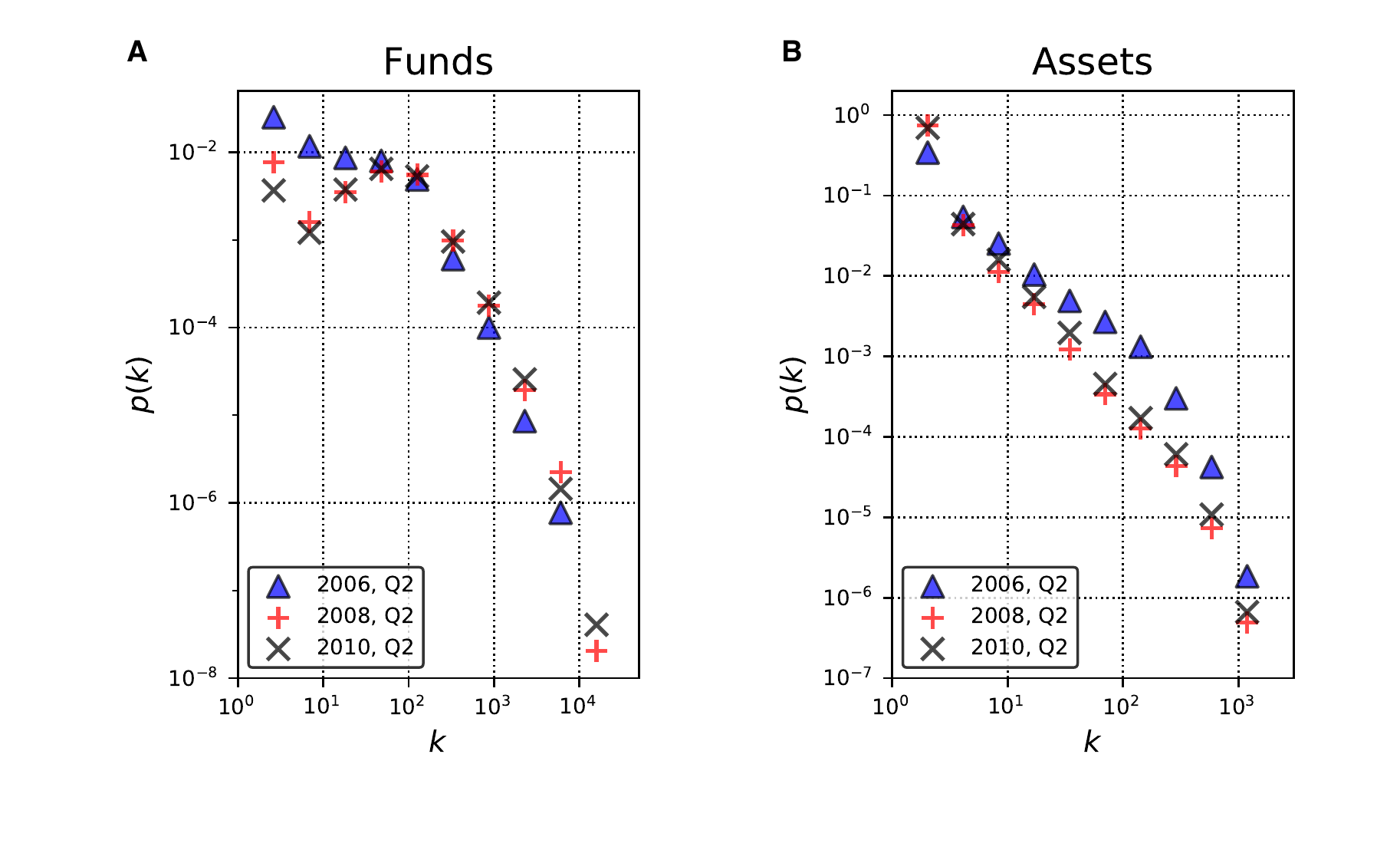}
  \caption{{\bf Degree distributions.}
  Probability density functions of fund degree (\textbf{A}) and asset degree (\textbf{B}) for the three reference quarters.
  Most funds invest in few tens of different assets but some manage very diversified portfolios, with thousands of different holdings.
  The assets present similar heterogeneity. During the crisis, the probability of holding many assets has increased, while the probability for an asset 
  to be held by many funds has fallen.
  }
  \label{fig:deg}
\end{figure}

The probability distributions $p(k_{i,\alpha})$ for the fund and asset degree, respectively, are shown in Fig.~\ref{fig:deg} for the selected reference quarters.
The probability densities extend over several orders of magnitude in both cases, with fat tails that are the typical signature of topologically 
heterogeneous systems.
This means that most funds manage portfolios of just few tens of different assets while the average degree is in the hundreds. However, there are funds 
for which the number of assets is larger than the average by orders of magnitude (hubs).
In between these two extremes we find a wide spectrum of intermediate investment strategies.
A similar heterogeneity characterizes the assets: most assets are found in the portfolios of few funds, but some assets enjoy huge popularity and are held by almost every fund.

During crisis, the tail of the distribution for funds has grown ``fatter'', while an opposite behavior is observed for assets.
We reckon that, during crisis, the probability of finding portfolios with a small degree has decreased and, correspondingly, the probability for a fund to invest in a large number of different financial products has increased.
At the same time, the probability of finding the same asset in many portfolios has become smaller, suggesting that the average overlap between portfolios has decreased.
\subsubsection*{Fund diversification}
Fund-degree provides a first topological notion of diversification of an individual portfolio.
However, from a financial point of view we must take into account the weights of each asset in a portfolio and consider a specialized index of concentration.

The Herfindahl--Hirschman index is a typical measure of concentration~\cite{Battiston:2004aa,Glattfelder:2009aa,Woerheide:1993aa} and we use its 
inverse to compare the diversification of mutual fund portfolios.
Precisely, we consider the following index
\begin{equation}
  h_i = \left(\sum_{\alpha} w_{i\alpha}^2\right)^{-1}\,.
  \label{eq:h}
\end{equation}
If a fund manager invests primarily in a single asset, $h_i$ will be
close to one; the opposite case is represented by a uniform investment, with
the same fraction of portfolio wealth invested in each asset, when we have $h_i
= k_i$. In this sense, $h_i$ can be regarded as the number of leading assets in the portfolio.
We will refer to the quantity defined in Eq.~\eqref{eq:h} as the ``inverse Herfindahl index''.

The larger the value of $h$, the more the portfolio can be considered as diversified.
This notion of diversification does not take into account how similar are different portfolios.
Even though diversification is believed to lower market risk in a portfolio, we will show that a systemic risk component arises since portfolios 
tend to diversify in the same way.
This adds a second systemic dimension to the notion of diversification, that is of major interest in a network perspective.
\subsubsection*{Portfolio similarity}
Portfolio similarity depends indeed on two factors: the number of assets they have in common, and the similarity of the weights attached to common assets.
Given two funds $i$ and $j$, let $P_i$ and $P_j$ be the sets of the indices of the assets in their portfolios, that is $P_i=\{\alpha_l^{(i)}\}_{l=1,\dots,k_i}$ 
and $P_j=\{\alpha_l^{(j)}\}_{l=1,\dots,k_j}$, with $\alpha_l\in [1,N_a]$.
We indicate as $|P|$ the cardinality of the set of assets $P$.
A measure of overlap between the various choices is provided by the \emph{Jaccard Index}~\cite{Jaccard:1901aa}, which is the 
size of the intersection of the two sets divided by the size of their union:
\begin{equation}
  J_{ij} = \frac{|P_i \cap P_j |}{| P_i \cup P_j |}\,.
  \label{eq:jaccard}
\end{equation}
This shows which percentage of the assets included in $P_i$ or $P_j$ belongs to
both. When two portfolios contain exactly the same assets, the Jaccard Index equals one, $J=1$.

We notice that asset degree is connected to the notion of overlap between portfolios.
Indeed, the larger the degree of an asset, the more ``popular'' it is in investment strategies and the larger the probability of finding it in different portfolios.

The Jaccard index refers to a topological notion of similarity but portfolio weights must be taken into account to provide a more relevant measure.
Indeed, when two funds invest in the same assets, their portfolios should be considered dissimilar in financial terms if the distribution of the weights are different.
To capture this, we sum the smallest of the weights attached to each asset in the intersection and multiply this by the Jaccard index, obtaining a \emph{Similarity Index}~\cite{Wool:2013aa}.
Formally, the latter can be defined as follows:
\begin{equation}
  s_{ij} = J_{ij} \times
  \sum_{\alpha \in P_i \cap P_j} \mathrm{min}\left( w_{i\alpha},w_{j\alpha} \right)\,.
  \label{eq:SI}
\end{equation}
This index is now equal to one for two portfolios that contain exactly the same assets
in exactly the same proportions, and it will be smaller than one otherwise.
When the two portfolios do not overlap at all its value is zero.

\begin{figure}[!t]
  \centering
  \includegraphics[width=1.\textwidth]{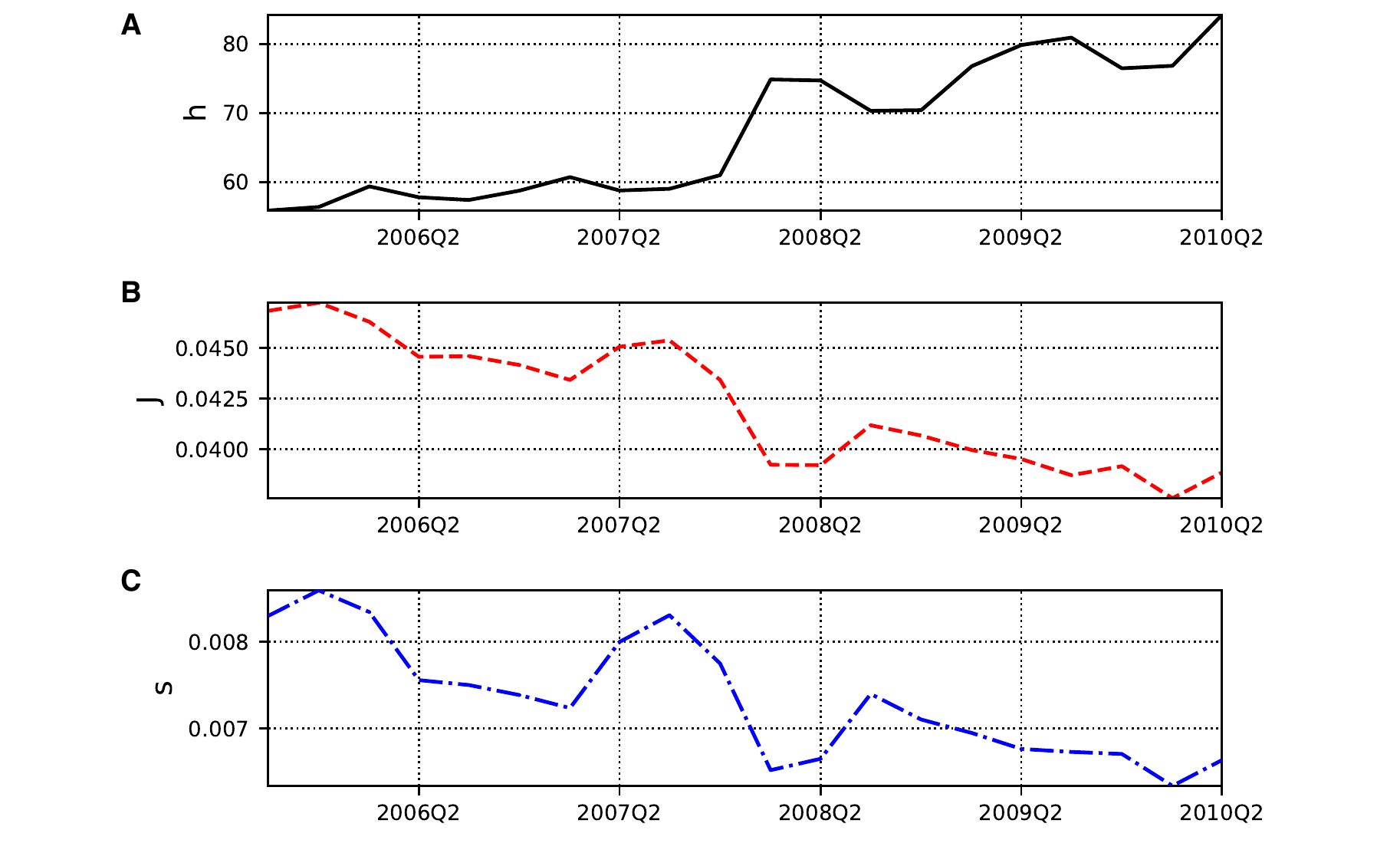}
  \caption{{\bf Portfolio diversification and similarity over time.}
  Average values of Herfindahl index (\textbf{A}), Jaccard index (\textbf{B}) and similarity index (\textbf{C}). 
  During the global financial crisis, the level of diversification within portfolios has increased while the level of similarity across portfolios has reduced.}
  \label{fig:diversification_tl}
\end{figure}

In Fig.~\ref{fig:diversification_tl} we show the level of concentration and similarity in the network over time, 
measured by the average of the indicators introduced before.
A considerable increase of diversification has taken place during the crisis, with an increment of 27\% in the h-index in 2007Q2--2008Q2.
Conversely, the average amount of overlap between two portfolios has reduced, with percentage variations of -13\% and -17\% for the Jaccard index and the 
similarity index respectively.
These results support the scenario outlined from the analysis of degree distributions: during crisis, fund managers raised the level of diversification within their portfolios; at the same time, investment decisions of individual managers are more ``diversified'' from each other.
This picture is also in good agreement with trends in the basic quantities that define the topology of the network (see the summary statistics reported in Table~\ref{S1_Table} in the Supplementary Materials section).
During the crisis the average degree of funds has nearly doubled.
More evidently, the total number of assets in the network has grown by an order of magnitude and the network density has reduced correspondingly. The growth in the number of assets in mutual funds during the crisis has been recently documented also by using other data sources, such as the Pensions and Investments survey~\cite{Sialm:2015aa}.
It is worth noting that the enlargement of the set of different assets is crucial for understanding the opposite trends of diversification and similarity in Fig.~\ref{fig:diversification_tl}.
Indeed, if the number of available asset stayed the same, we would expect an increase of diversification of the portfolios to be reflected, to some extent, in an increase of their overlap.

From Fig.~\ref{fig:diversification_tl} we can also observe that the timeline of the similarity $s$ closely mimics the Jaccard index $J$.
This suggests that, on average, the similarity between two portfolios in the holding network is mainly determined by the commonality of their assets.
This was only partly predictable.
Indeed, conditionally on the weights $w_{i,\alpha}$, the value of $s$ is proportional to $J$ (see Eq.~\eqref{eq:SI}).
However, in principle the two indicators could have different behaviors and we could imagine cases where, for instance, the relative 
size of the intersection of two portfolio increases but the value of $s$ decreases if the weights attached to common assets are tuned suitably.

\subsection*{Comparison to null models}
We analyse the level of heterogeneity and information content in the network of holdings by comparing the structure of the real network with 
two different ``null'' models. These models are obtained through randomization of the original investments by two alternative strategies.

In the first case, we start from a graph with $N_f$ funds, $N_a$ assets and no links, and then we reassign the original $E$ edges (with their 
original market values attached) drawing at random and without repetition from the possible fund--asset pairs.
Naively, we can think of this null model as a representation of a scenario where portfolio managers choose at random in which assets to invest and 
how much money to put on each of them. In the text and the figures we will identify this null model with the abbreviation \emph{Rnd-1}.

In the other case, we perform the randomization subjected to the constraint that the degree and portfolio weights of each fund do not change.
This is accomplished by reassigning the original investments of each fund to randomly selected assets (by construction, this strategy preserves 
the original value of the Herfindahl index of portfolios). This model represents a network where fund managers set the level of diversification 
in advance; that is, they decide the number of distinct assets to invest in, as well as the relative proportions, and they then choose the assets at random. 
We indicate in the following with the abbreviation \emph{Rnd-2} the second null model.
Since values of portfolio concentration in this model are the same as in the original network, we shall exploit it to separate 
the contribution of similarity to systemic fragility. In summary, while the number of edges is preserved in the both cases,  the degree 
sequence of funds is preserved only in the second one.

\begin{figure}[!t]
  \centering
  \includegraphics[width=1.\textwidth]{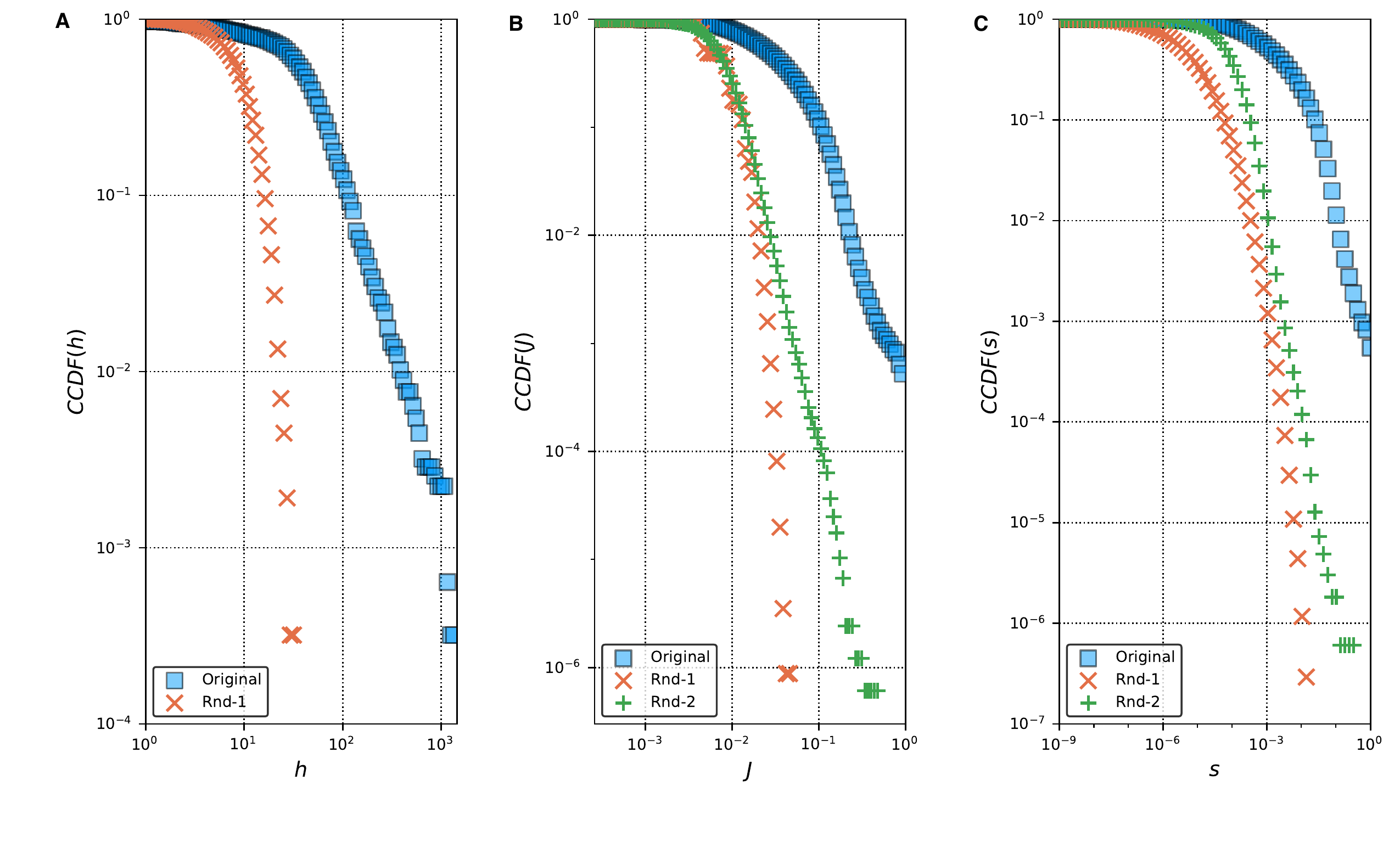}
  \caption{{\bf Diversification and similarity: comparison to the null models.}
  Complementary cumulative distribution functions of Herfindahl index, $h$, (\textbf{A}), Jaccard index, $J$ (\textbf{B}) and similarity index, $s$ (\textbf{C}).
  The real network in 2006Q2 is compared to the null models as described in the text.
  The latter ones fail in reproducing the heterogeneous behaviour observed in the real systems.
  In particular, large values of the number of effective assets ($h$) and of the similarity between portfolios ($s$) are much more likely in the 
  real world than in the randomised sets.}
  \label{fig:divdists}
\end{figure}

In Fig.~\ref{fig:divdists} we compare the complementary cumulative distribution functions ($CCDF$) of the  different indices ($h,J,s$) in a snapshot of the real network (2006Q2) compared with their distribution measured in the two null models.
The $CCDF(x)$ gives the probability of finding a value of the index that is greater or equal to $x$.
The same analysis done in the other quarters provides similar results.
In particular, we find that the distribution of $h$ provides a clear picture of the system's heterogeneity.
In the real network, $p(h)$ decays slowly and the number of effective assets in portfolios ranges from few to thousands.
Conversely, the distribution for model Rnd-1 is concentrated and the probabilities of large $h$ are smaller by orders of magnitude.
In the real case, the average value of leading assets in a portfolio is $\bar{h}\approx 58$, while it is only $9$ for model Rnd-1.
Similar considerations apply when we inspect the probability distributions of $J$ and $s$.
In these cases, a comparison is also performed with model Rnd-2, which retains the same values of diversification of the original network.
We see that large values of the indices are more likely to occur for model Rnd-2 than Rnd-1, but both models fail to capture the probability of very large overlap that characterizes the real network.
In particular, only in the real case we find pairs of almost identical portfolios.
As a matter of fact, community detection on the network reveals the tendency of funds to cluster in small groups of portfolios with very similar structures.

As far as diversification of individual portfolios is concerned, the result of the comparison supports intuition.
Fund managers are professional investors whose diversification strategies cannot be reduced to random selection of assets.
As for the overlap of different portfolios, we conclude that large similarities are considerably more likely than could be expected by chance.
This result still holds when the values of $h$ of real portfolios are kept fixed during the randomization procedure.
This indicates that the effects of diversification \emph{per se} are not sufficient to explain the probability of large similarity observed in the 
network and suggests that investment strategies of different funds can be strongly correlated. Many different mechanisms have been suggested in the literature to
account for such a high degree of similarity across portfolios including connections between mutual fund managers and corporate board members, herding behavior and imitation
of successful diversification strategies.
We will show that such correlation between portfolios contributes to a large extent to the system's vulnerability to financial shocks.

\subsection*{Systemic risk and similarity}
To test the effects of similarity on the network fragility, we consider a basic model of distress propagation.
An idiosyncratic shock in the value of an asset spreads across the network due to common exposures of portfolios (more details are provided in Materials and Methods).
The shock propagates because individual investors sell their portfolio shares in response to a drop in the value of the fund they invest in.
Selling shares entails selling the individual assets in the portfolio.
This process, produces a negative effect on the market value of assets being sold and triggers a new round of losses for the portfolios owning those assets.
This is, of course, a simplified view of the network dynamics which does not consider portfolio reallocations. However, reallocation of mutual fund investments
is a slow process which might only partially ameliorate the negative impact of the asset sell-off process.

While being an approximation, the assumption of a static network can be justified with the following arguments.
First, we expect portfolio reallocations by fund managers to take place on a time scale larger than the one of asset price movements on the market 
and of buying/selling decisions of individual investors.
Moreover, keeping the topology fixed, we can provide a simple measure of fragility for each snapshot of the network and study its evolution during crisis.

In the context of such dynamics, we measure network fragility by considering the total percentage ``loss of value'' of the system.
Since the total loss depends strongly on the asset receiving the initial shock, we compute the average $D$ of the losses induced by a shock to one of the 
``top assets''. We define the latter as the assets whose market value in the portfolios of all funds is larger than the percentile of order 0.999 
of the distribution.
The quantity $D$ provides a measure of the \emph{systemic damage} produced by a single shock, as a function of the time elapsed and of the shock entity.

\begin{figure}[!t]
  \centering
  \includegraphics[width=1.0\textwidth]{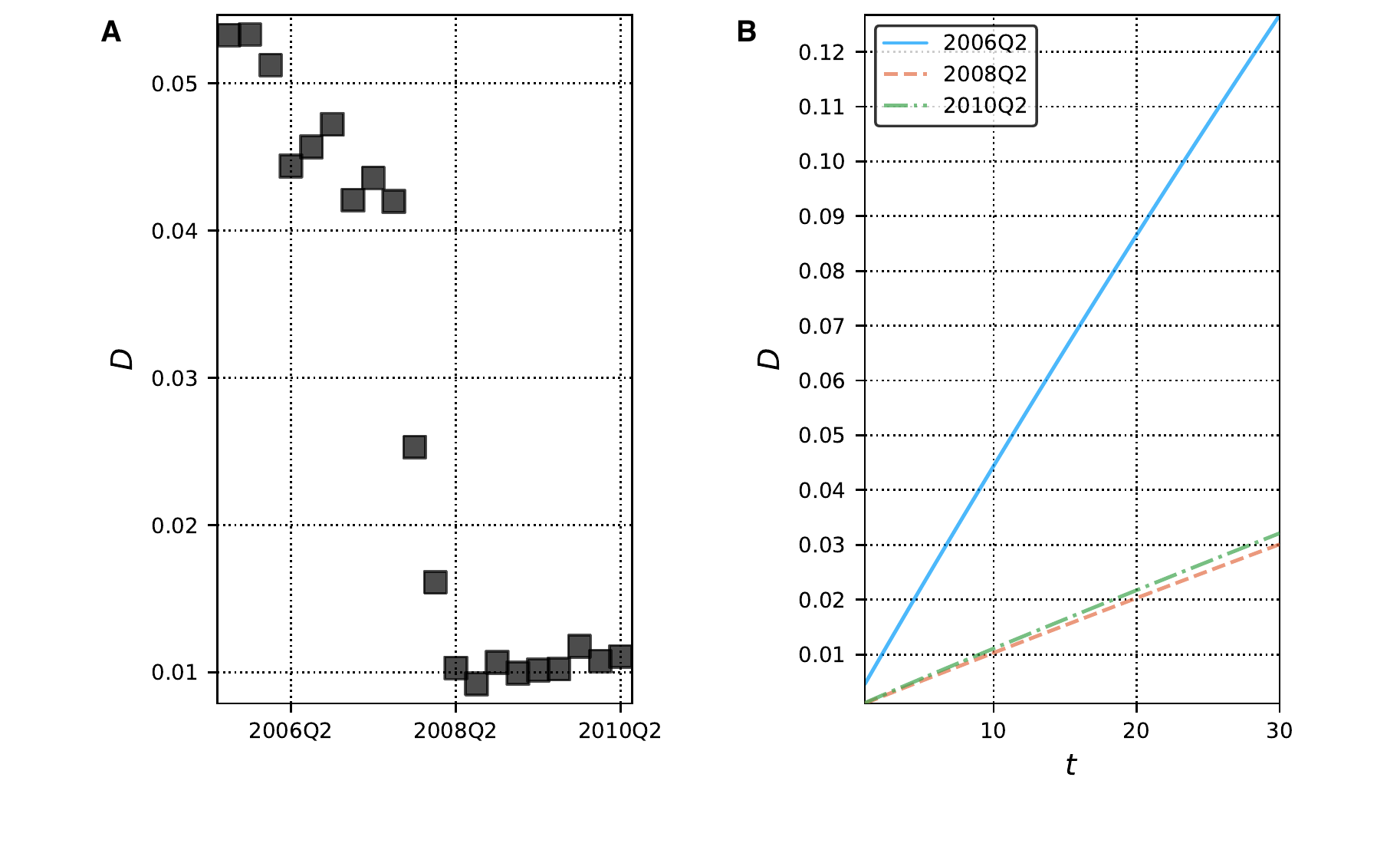}
  \caption{\textbf{Shock propagation before and after the global financial crisis.}
    Systemic fragility has fallen during the global financial crisis \textbf{(A)}.
    Before the crisis, a single shock would have provoked a considerable loss for the whole system after few time steps \textbf{(B)}.
    The velocity of damage propagation after the crisis is \emph{1/4} than before but a systemic vulnerability still exists.
    Both panels correspond to the results of an initial shock $\delta_0=50\%$; panel \textbf{A} reports the damage after $t=10$ steps.}
  \label{fig:damage_evo}
\end{figure}

Previously, we have shown that the average similarity between portfolios has decreased during crisis, see Fig.~\ref{fig:diversification_tl}.
We observe from Fig.~\ref{fig:damage_evo} that  similarity is correlated with fragility, since the network has indeed become more robust.
The potential systemic damage in the network has been progressively reduced in the years 2005--2010 (panel \textbf{A}).
Before the crisis, a 50\% shock to just one of the top assets would give a 4–6\% loss in the whole system's value after 10 time steps.
To put this into perspective, such a loss would correspond to 120–180 billion dollars in currency amount in 2006Q2.
The loss reduces to 1\% after crisis.
As already observed for the average similarity, this trend has already started at the beginning of the observation period but has 
undergone a strong acceleration during the global financial crisis.
Moreover, before the GFC distress propagation was a fast process (panel \textbf{B}) and damage could hit the 10\% barrier in few steps.
After the crisis the velocity of damage propagation, estimated by the slope of $D(t)$, is four times smaller.
Nonetheless, a vulnerability still exists.

\begin{figure}[!t]
  \centering
  \includegraphics[width=1.0\textwidth]{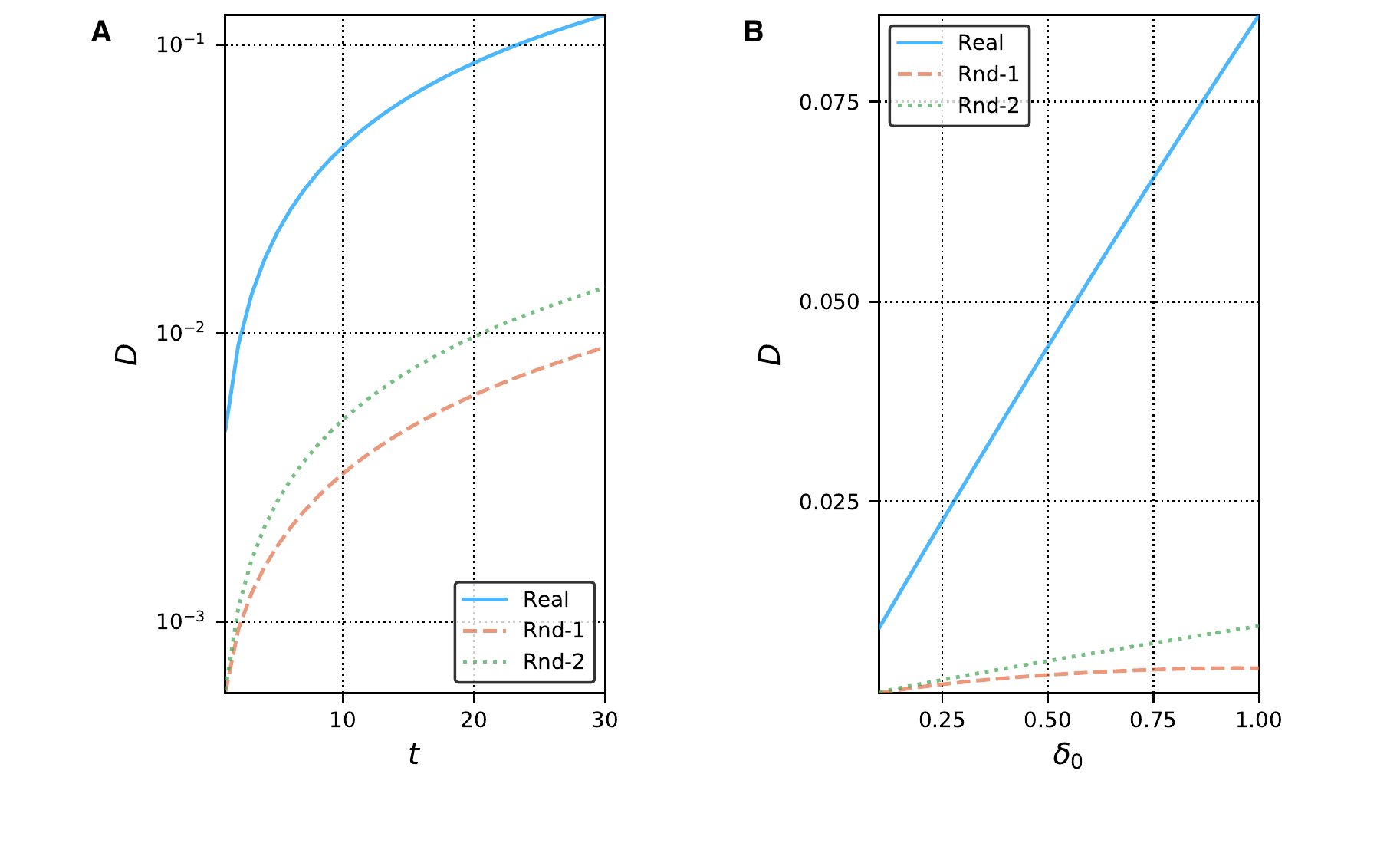}
  \caption{\textbf{Effect of similarity on systemic fragility.}
    Values of the systemic damage as a function of time for $\delta_0=0.5$ \textbf{(A)} and as a function of shock entity for $t=10$ \textbf{(B)}.
    Randomization has a positive impact: damage propagation in the real case is 8.8 times faster than for model Rnd-2 and 14.5 time faster than model Rnd-1.
    After randomization, the propagation is also less sensitive to shock entity.
    Comparison of models Rnd-1 and Rnd-2 indicates that systemic fragility can be ascribed mainly to the similarity between portfolios and that systemic damage is minimum for completely random investments (Rnd-1).
    }
  \label{fig:damage_sensitivity}
\end{figure}

In the two panels of Fig.~\ref{fig:damage_sensitivity} we compare the systemic damage in a sample snapshot (2006Q2) of the real system 
with its random counterparts.

The comparison is performed for a fixed value of the shock ($\delta_0=0.5$) as a function of time, and for a fixed time ($t=10$) as a function 
of the shock, respectively.

The result is that systemic fragility reduces by at least an order of magnitude when a randomization is performed.
In model Rnd-1, where the diversification profile of the original portfolios is altered (we obtain a model of random investments), the values of $h$ are concentrated, the average is smaller that in the original network, and large values of $h$ are simply not observed (see Fig.~\ref{fig:divdists}).
Furthermore, the overlap between portfolios is also largely suppressed.
In model Rnd-2, the original values of $h$ are preserved and only the similarity is reduced since assets are reassigned randomly.
From that property follows that the reduction of systemic damage that we observe for model Rnd-2 can be considered a direct effect of the reduction in the similarities between portfolios.
The damage for Rnd-1 is even smaller, the difference getting larger with both time and shock entity.
Counter-intuitively, a model of random investments, where portfolios tend to have similar diversification and a small number of leading assets, is more robust with respect to the simulated dynamics of shock propagation.

We conclude that significant similarities between portfolios, a statistical signature of the holding network, is crucial in 
the transmission of financial distress and can make the network more fragile.
Moreover, in a network of random investments (Rnd-1) this systemic risk component is reduced further and the network appears 
less fragile even though portfolios are more concentrated.

\section*{Discussion}
We perform the first extensive study of the structure and the evolution of the US mutual fund network throughout the Global Financial Crisis (GCF) of 2007--2008. Even though in normal times households rarely rebalance their retirement saving portfolios~\cite{Agnew:2003aa}, it has been found that 21\% of them changed investment strategy between February and November 2009~\cite{Hurd:2010aa}. Such a dramatic recomposition of investment portfolios during crises can have severe consequences on financial stability.
The size of the ``ecosystem'' of different fund investments has grown steeply over time~\cite{Sialm:2015aa} and, as an average, in the aftermath of the crisis mutual funds have become better diversified and less similar. However, mean values do not tell the whole story.
Inspection of the probability density functions shows evidence of an heterogeneous system, with few largely diversified hubs and many specialized funds.
Moreover, the probability of the similarity between portfolios decays slowly and large similarities are far more likely that can be expected from 
benchmark models of a random network of investments. We conclude that a high degree of correlation exists between investment decisions of different funds.
This correlation limits the effectiveness of fund of funds diversification strategies.

One of the leading forces behind the emergence of such correlation can be found in the social network of relationships between fund managers~\cite{Augustiani:2015aa} and the effects of managerial sharing.
Other reasons may be due to herding behavior and the fact that professional investors with similar targets and risk profiles are likely to adopt similar investment strategies.
For instance, portfolio managers try to maximize profits and the strategies of many of them will likely include those assets that have proved to be profitable or that can be selected by shared quantitative analysis techniques. 
Extreme market uncertainty can act as a driver for fund investments during a crisis, when an important fraction of their invested capital is moved from equity mutual funds to fixed-income mutual funds. During the crisis, defined contribution equity mutual funds experienced a large outflow of more that -15\%, while flow into the fixed income mutual funds reached an historical peak of +20\% \cite{Sialm:2015aa}.
Similarly, many funds might be damaged during crisis at the same time and trigger a second-order effect by which other funds get 
hit in a failure cascade.
In our stylized representation of distress propagation, such second-order effect is induced by the many individual investors that simultaneously sell fund shares.
However, a complete representation would also consider changes in the network topology as a result of the fact that portfolio managers will 
try to rebalance their portfolios.
Massive co-movements in fund allocations as a response to crisis may have an even higher impact on the market value of securities.
This, in turn, may result in significant effects back to the mutual fund network and possibly lead to higher levels of overlap between funds.

Similarity of portfolios provides a different notion of diversification of investments.
Exploiting a stylized model of shock propagation on a static network, we have shown how strongly this sort of systemic coupling can impact the 
fragility of the network.
We find that the systemic damage in the network is much larger than the damage that can be procured to a network of random investments. Thus we quantify
the systemic risk induced by the similarity of portfolio investment strategies.
The comparison between a completely random model and a model where the diversification of original portfolios is preserved shows that similarity 
between portfolios represents a major source of systemic risk that can make the network fragile.
It also emerges that completely random investments, even though corresponding to more concentrated portfolios, would result in a less vulnerable network.
That is not to say that diversification is self-defeating \emph{per se}, but indicates that it can make the system more fragile as funds tend to 
diversify their portfolios with similar investment patterns. 
The global financial crisis has stimulated an increase in the diversification of portfolios, but a systemic risk component still exists because of the similarity of investments.

We believe that the evidences presented in our study have implications for both the modeling and the regulation of financial networks.
They show that similarity induces a systemic vulnerability and that portfolio diversification may even play antagonistically when strong correlations 
between investment strategies exist.
Both aspects should be taken into account for the purpose of assessing systemic risk in holding networks and
for devising effective policy actions.


\section*{Materials and Methods}
%
\paragraph{Dataset} The US Mutual Fund market is the largest in the world. With 15\$ trillion assets under management at year-end 2013, it accounts for 
about half the total value in mutual fund assets worldwide.
In this study, we analyse data from the CRSP Survivor-Bias-Free Mutual Fund Database. 
It includes open-ended mutual funds registered with the Securities and Exchange Commission and provides detailed information about the composition of 
managed portfolios.
A mapping of funds to the portfolios of assets they manage is provided and detailed information about portfolio holdings, including the market value 
of each holding, is delivered on a quarterly basis.

\begin{figure}[!t]
  \centering
  \includegraphics[scale=.9]{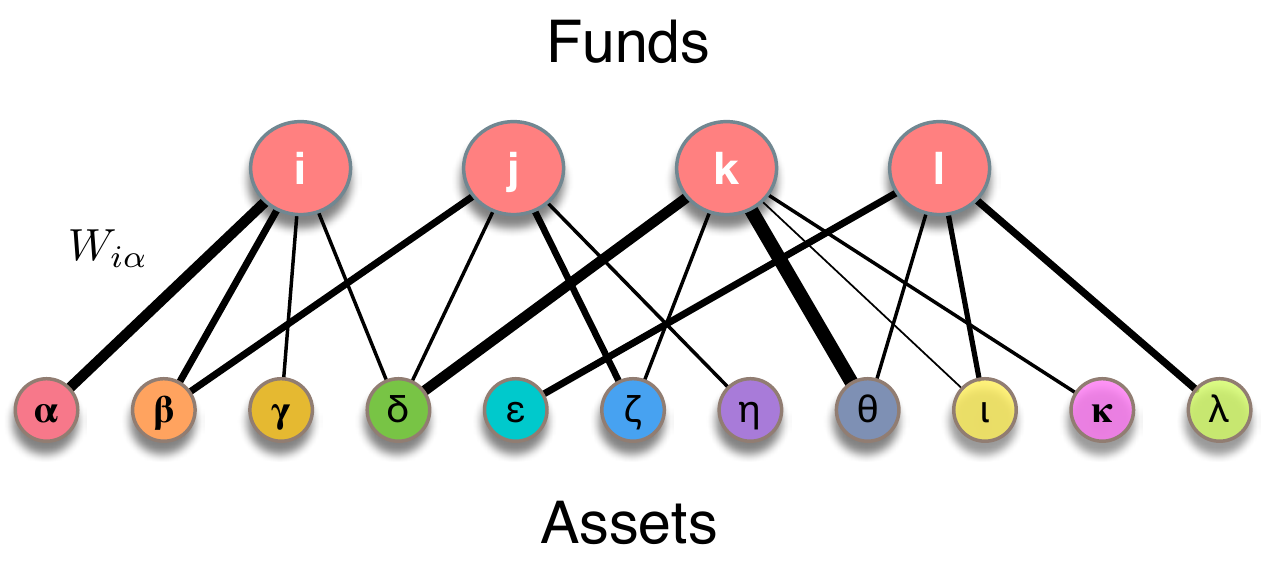}
  \caption{{\bf Graph of portfolio holdings.}
  The network of portfolio holdings can be represented as a bipartite graph.
  The two vertex classes are the \emph{funds} $\{i,j,k,\dots\}$ and the \emph{assets} $\{\alpha,\beta,\gamma,\dots\}$ in their portfolios.
  Each edge $(i,\alpha)$ represents a specific holding relationship.
  The edge weight $W_{i\alpha}$ is equal to the total market value of security $\alpha$ owned by fund $i$ in its portfolio.}
\label{fig:bptgraph}
\end{figure}

In the following, we provide a formal construction of the bipartite network of holdings.
We also present a schematic description of the model of distress propagation considered in the main text.
\paragraph{Bipartite graph of portfolio holdings}
We represent portfolio holdings in terms of a bipartite graph.
The two vertex classes are the set of mutual funds $i=1,\dots, N_f$, and the set of the different assets $\alpha=1,\dots, N_{a}$ in their portfolios.
The degree $k_i$ of vertex $i$ is exactly the number of distinct assets held in the portfolio of fund $i$.
Edges incident at vertex $i$ can be assigned weights $W_{i\alpha}$ equal to the total market value of the shares of asset $\alpha$ and the graph is also 
conveniently represented by a $N_f \times N_a$ matrix $B$ with elements $B_{i\alpha}=W_{i\alpha}$.
We indicate as $G(N_f, N_a)$, or simply $G$, the undirected bipartite network of portfolio holdings.
If we wanted to retain just the topological information of which asset is owned by which fund, we could define an unweighted graph $G^0(N_f, N_a)$.
This would correspond~\cite{Caldarelli:2007aa,Newman:2010aa} to the incidence matrix $B^0$ whose elements $B_{i\alpha}^0$ are $1$ if asset $\alpha$ is 
in the portfolio of fund $i$ and $0$ otherwise.
A schematic picture of the network of holdings is provided in Fig.~\ref{fig:bptgraph}.

The \emph{size} of fund $i$ is given by
\begin{equation*}
  S_{i}= \sum_{\alpha=1}^{N_a} W_{i\alpha}\,,
\end{equation*}
where it is understood that $w_{i\alpha}=0$ if asset $\alpha$ is not in the portfolio.
The quantities $W_{i\alpha}$
We easily see that, in network terms, fund size corresponds to \emph{node strength}~\cite{Barrat:2004aa,Yook:2002aa}.
We also indicate as $M_{\alpha}$ the aggregate market value of an asset
\begin{equation*}
  M_{\alpha} = \sum_{i=1}^{N_f} W_{i\alpha}\,.
\end{equation*}
The sum $S_{tot} = \sum_i S_i$ represents the \emph{total value} of the system.
By construction it holds $\sum_{i=1}^{N_f} S_i = \sum_{\alpha=1}^{N_\alpha} M_{\alpha}$.

The quantities $S_i$ and $W_{i\alpha}$ are expressed in currency units.
We also define the \emph{portfolio weights} $w_{i\alpha}=W_{i\alpha}/S_i$, that represent the fraction of portfolio wealth corresponding to each asset.
The indices of diversification and similarity discussed in the main text are all expressed in terms of the $w_{i\alpha}$s.

Due to portfolio reallocations, the set of assets in a portfolio and the edge weights can change with time; also the number of indexed funds in the dataset depends on time.
Reporting of portfolio reallocations by funds are not synchronous and some choices are due in order to aggregate the information about holdings over some time horizon and construct the graph representing the system.
Since portfolio composition is reported on a quarterly basis, in our analysis we choose a three-month time frame for aggregation.
We create quarterly snapshots of the bipartite network by means of the following procedure.
For each quarter, we consider the set of the funds $i=1,\dots, N_f$ for which a mapping to a portfolio exists.
For the given quarter and for each fund $i$, we retrieve the holding information at the most recent report date $t_i$.
Basically, this information is given by the set of assets in the portfolio and the corresponding market values.
Fund $i$ is inserted into the graph for the current quarter and so are the links $(i,\alpha)$, with $\alpha\in [1,N_a]$, for all assets held by the fund.
Each link is assigned weight $W_{i\alpha}$ equal to the market value of the holding it represents.

When parsing the holding relationships in the dataset, attention has to be paid to the issue of \emph{fund classes}.
As a matter of fact, a fund may issue different types of shares all corresponding to the same underlying portfolio.
In the database, different classes of a funds are associated to different unique identifiers, as if they were distinct funds.
For the purpose of our analysis, we consolidated information about fund classes, to avoid including the same fund multiple times when constructing 
the network snapshots.

\paragraph{Basic model of distress propagation}
Let us suppose that at time $t$ the prices of some securities undergo a negative shock $\delta_\alpha(t)=[v_\alpha(t+1)-v_\alpha(t)]/v_\alpha(t)\leq 0$, 
where $v_{\alpha}(t)$ is the market price of $\alpha$.
These downward moves produce a negative variation in the values of the portfolios that hold the securities in object.
As a consequence, we expect individual fund investors to disinvest their shares, effectively selling the individual securities of the corresponding portfolio.

Generally, we expect that to a larger drop $\Delta_i$ of the portfolio of fund $i$, corresponds a higher probability $p(\Delta_i)$ for an investor 
to disinvest its shares of the fund.
Let $V_{\alpha}$ be the total market value of $\alpha$ disinvested in the process of selling fund shares.
The instantaneous increase in the offer of $\alpha$ will have a negative impact in its value on the market.
This reflects in a new drop $\delta_\alpha(t+1) = \lambda(V_\alpha(t))$, where $\lambda$ is some \emph{market impact} function (deterministic or stochastic).
In the absence of large random effects and portfolio reallocations, in a connected network of portfolio holdings such a recursive dynamics will 
bring a progressive reduction of the total value of the system.

For the purpose of simplicity and clarity, the simulations discussed in the main text assume the fraction of fund shares liquidated to be equal to the portfolio drop and the market impact to be proportional to $V_{\alpha}$.
Such a model provides a simplified, schematic representation of shock propagation within a bipartite network of portfolios and their holdings.
It should be considered an approximation of the dynamics in the time span that divides subsequent reallocations executed by fund managers, wherein 
the network can be regarded as static.
It incorporates \emph{contagion effects} due to common exposures of overlapping portfolios and is a multiplicative process where initial shocks get 
amplified over time.

To compare the systemic fragility of the network at different times, we performed simulations where a single security $\alpha$ is given an initial shock.
This shock propagates for a number of steps and at each time the percentage total loss of value of the system is computed.
So, the damage at time $t$ given a shock to $\alpha$ is computed as $D_\alpha(t)=(S_{tot}(t+1)-S_{tot}(0))/S_{tot}(0)$.
The process is repeated for a given set $A$ of securities and we define the \emph{systemic damage} by the average $D(t)=1/|A| \sum_{\alpha\in A} D_\alpha(t)$.
In order to reduce the computational time and speed up the process, we choose $A$ as the set of assets for which the total value in the portfolios of funds is individually above the quantile of order 0.999 (\emph{top assets}).\\[37pt]

\subsection*{List of Supplementary Materials items}
\begin{itemize}
  \item[-] Table~\ref{S1_Table}: \textbf{Summary statistics of the bipartite network of holdings.}
\end{itemize}

\newpage

\noindent \textbf{Acknowledgements:} 
Guido Caldarelli acknowledges support from EU projects Multiplex 317532, Simpol
610704, Dolfins 640772, CoeGSS 676547 and SoBigdata 654024.

\noindent \textbf{Competing Interests} The authors declare that they have no
competing financial interests.\\


\clearpage


\section*{Supplementary Materials}
\renewcommand\thetable{S\arabic{table}}
\setcounter{table}{0}

\renewcommand\thefigure{S\arabic{figure}}
\setcounter{figure}{0}

\begin{table}[!h]
  \centering
  \begin{tabular}{lccccc}
    Quarter & $N_f$ & $N_a$ & $\rho$ & $\bar{k}$ & $S_{tot}$ (\$B) \\
    \midrule
    2005Q3 & 2474 & 7498 & 0.015 & 109 & 2123.7\\
    2005Q4 & 2233 & 7117 & 0.016 & 113 & 2523.8\\
    2006Q1 & 2545 & 7810 & 0.015 & 121 & 2767.7\\
    2006Q2 & 3128 & 10743 & 0.011 & 114 & 2944.9\\
    2006Q3 & 3197 & 10878 & 0.010 & 112 & 2980.4\\
    2006Q4 & 3186 & 11471 & 0.010 & 117 & 3221.5\\
    2007Q1 & 3155 & 12410 & 0.010 & 123 & 3042.2\\
    2007Q2 & 3609 & 12482 & 0.010 & 121 & 3738.0\\
    2007Q3 & 3492 & 16303 & 0.008 & 128 & 3848.2\\
    2007Q4 & 3414 & 37694 & 0.004 & 144 & 4319.9\\
    2008Q1 & 3773 & 74875 & 0.003 & 189 & 4983.3\\
    2008Q2 & 5007 & 158894 & 0.001 & 198 & 5617.8\\
    2008Q3 & 5411 & 197897 & 0.001 & 194 & 4626.6\\
    2008Q4 & 5453 & 136511 & 0.001 & 189 & 3718.1\\
    2009Q1 & 5689 & 164507 & 0.001 & 200 & 4653.8\\
    2009Q2 & 5733 & 152444 & 0.001 & 203 & 5355.9\\
    2009Q3 & 5689 & 147706 & 0.001 & 205 & 6066.5\\
    2009Q4 & 5751 & 119412 & 0.002 & 182 & 6036.0\\
    2010Q1 & 5511 & 131137 & 0.001 & 187 & 6300.3\\
    2010Q2 & 5022 & 122458 & 0.002 & 202 & 5957.7\\
    \bottomrule
  \end{tabular}
  \caption{
    {\bf Summary statistics of the bipartite network of holdings.}
    Number of funds $N_f$ and assets $N_a$, the network density $\rho$, the average number of assets in portfolios $\bar{k}$ and the total value of 
    the network expressed in billion dollars.
    Network density is defined as $\rho=E/(N_f N_a)$ where $E$ is the total number of investments (number of portfolio–asset holding relationships).
    The number of indexed funds and different assets in their investments have grown systematically.
    The average degree of funds nearly doubles during crisis and the network becomes less dense, with $\rho$ decreasing by an order of magnitude.
  }
  \label{S1_Table}
\end{table}

\end{document}